\documentclass[reprint,amsmath,amssymb,aps,superscriptaddress,floatfix]{revtex4-1}
\usepackage{color}
\usepackage{amsmath}
\usepackage{graphicx}
\usepackage{dcolumn}
\usepackage{bm}
\usepackage{booktabs}
\usepackage{multirow}
\usepackage[colorlinks]{hyperref}
\usepackage{lineno}

\hypersetup{
	linkcolor = {blue},
	citecolor = {blue},
	urlcolor = {blue}
}

\makeatletter

\newcommand{\Rmnum}[1]{\expandafter\@slowromancap\romannumeral #1@}
\makeatother

\begin{document}
\preprint{APS/123-QED}
	
\title{Regeneration of Spin Wave in Atomic Vapor}

\author{Jian-Peng Dou}
\affiliation{Center for Integrated Quantum Information Technologies (IQIT), School of Physics and Astronomy and State Key Laboratory of Advanced Optical Communication Systems and Networks, Shanghai Jiao Tong University, Shanghai 200240, China}
\affiliation{CAS Center for Excellence and Synergetic Innovation Center in Quantum Information and Quantum Physics, University of Science and Technology of China, Hefei, Anhui 230026, China}

\author{Feng Lu}
\affiliation{Center for Integrated Quantum Information Technologies (IQIT), School of Physics and Astronomy and State Key Laboratory of Advanced Optical Communication Systems and Networks, Shanghai Jiao Tong University, Shanghai 200240, China}
\affiliation{CAS Center for Excellence and Synergetic Innovation Center in Quantum Information and Quantum Physics, University of Science and Technology of China, Hefei, Anhui 230026, China}

\author{Xiao-Wen Shang}
\affiliation{Center for Integrated Quantum Information Technologies (IQIT), School of Physics and Astronomy and State Key Laboratory of Advanced Optical Communication Systems and Networks, Shanghai Jiao Tong University, Shanghai 200240, China}
\affiliation{CAS Center for Excellence and Synergetic Innovation Center in Quantum Information and Quantum Physics, University of Science and Technology of China, Hefei, Anhui 230026, China}

\author{Hao Tang}
\affiliation{Center for Integrated Quantum Information Technologies (IQIT), School of Physics and Astronomy and State Key Laboratory of Advanced Optical Communication Systems and Networks, Shanghai Jiao Tong University, Shanghai 200240, China}
\affiliation{CAS Center for Excellence and Synergetic Innovation Center in Quantum Information and Quantum Physics, University of Science and Technology of China, Hefei, Anhui 230026, China}

\author{Xian-Min Jin}
\thanks{xianmin.jin@sjtu.edu.cn}
\affiliation{Center for Integrated Quantum Information Technologies (IQIT), School of Physics and Astronomy and State Key Laboratory of Advanced Optical Communication Systems and Networks, Shanghai Jiao Tong University, Shanghai 200240, China}
\affiliation{CAS Center for Excellence and Synergetic Innovation Center in Quantum Information and Quantum Physics, University of Science and Technology of China, Hefei, Anhui 230026, China}
\affiliation{TuringQ Co., Ltd., Shanghai 200240, China}
\affiliation{Chip Hub for Integrated Photonics Xplore (CHIPX), Shanghai Jiao Tong University, Wuxi 214000, China}

\date{\today}	

\begin{abstract}
{Conventionally, atomic vapor is perceived as a non-living system governed by the principles of thermodynamics and statistical physics. However, the demarcation line between life and non-life appears to be less distinct than previously thought. In a study of amplifying spin waves, we observe a phenomenon reminiscent of life: The atomic spin wave stored in atomic vapor has a capability of absorbing energy from an external light source, and exhibits behaviors akin to active regeneration. We demonstrate that this regeneration significantly enhances the lifetime and retrieval efficiency of the spin wave, while concurrently the noise is effectively suppressed. Our results suggest that the regeneration mechanism holds promise for mitigating the pronounced decoherence typically encountered in spin waves carried by room-temperature media, therefore offering potential applications in the realms of quantum information and precision measurements at ambient conditions. }
\end{abstract}
\maketitle	

Regeneration in biology is a process of renewal, restoration, and tissue growth that enhances the resilience of organisms and ecosystems against natural fluctuations or events that cause disturbance or damage \cite{Carlson2007}. Similarly, if a fragile atomic spin wave possesses such a regenerative capability, its stability would be significantly enhanced, allowing it to maintain coherence and functionality in the face of disturbances or decoherence. Naturally, it's difficult to conceive that atomic spin waves would undergo an active regeneration akin to biological life without any empirical observation, since the carrier of spin wave, such as atomic vapor, is conventionally considered to be a non-living system governed by the principles of thermodynamics and statistical physics.

The motivation for regenerative spin waves arises from practical applications: Long-lived atomic spin waves, maintained within a quantum memory, form a pivotal cornerstone for large-scale quantum networks \cite{Wei2022, Lvovsky2009, Jing2019, Ladd2010, Zhang2023}. Furthermore, the related techniques also find applications in precision measurements, for instance, enhancing the measurement sensitivity of atomic magnetometers often requires optimizing spin-wave coherence lifetime \cite{Budker2007, Meng2023}. Moreover, spin waves within quantum memories have been instrumental in investigating fundamental physics, including multi-particle entanglement \cite{Li2021}, hybrid interference between light and atomic spin waves \cite{Shi2022}, and applications in space science \cite{Mol2023}.

Over the past two decades, the storage of atomic spin wave has been implemented in many quantum memory protocols as well as the corresponding storage media \cite{Jing2023, Davidson2023, Ming2023, Ma2022, Pu2021, Tang2015, Hosseini2009}. In order to improve the lifetime of spin wave, researchers have tried various protection methods such as ultra-low temperature, magnetic-field-insensitive spin wave \cite{Xu2013}, compensating differential light shift \cite{Yang2016}, dynamic decoupling \cite{Zhong2015}, spin orientation within a decoherence-free subspace of spin states \cite{Katz2018}, anti-relaxation coating and miniaturized vapor cell \cite{Dideriksen2021}. Among the above, those based on alkali-metal atomic vapor, as important candidates for realizing quantum memory, have attracted much attention due to low requirements on the working environment. The frequency splitting between the ground-state hyperfine energy levels of alkali-metal atom is large enough to distinguish signal photons and intense control light, such as 6.8\,GHz for $^{87}$Rb and 9.2\,GHz for $^{133}$Cs \cite{Steck2023}. This facilitates the frequency filtering for obtaining desired signal photons and large memory bandwidth, enabling quantum information to be processed at a high ‘clock rate’ \cite{Reim2010}. However, the corresponding spin-wave wavelength is limited to several centimetres, and the atomic motion and collisions lead to a fast decoherence of spin wave \cite{Finkelstein2023}, and the lifetime is of the order of microseconds usually determined by the time atoms take to escape from the region of spin wave.

For maintaining a long-lived atomic spin wave, besides the mentioned passive protection, a more straightforward way should have been the one: compensate the decay of spin wave by amplifying it actively, just like what an optical amplifier does for a weak light. It is worth noting that an amplification process is accompanied by intrinsic noise determined by quantum mechanics, especially Heisenberg’s uncertainty principle, which is well known in the research field of optical amplifier \cite{Inoue2018}. With this in mind, we embarked on investigating a kind of low-noise amplification of spin waves based on a fact that the spin wave is carried by a large number of atoms, and expected that the noise may be suppressed by some collective effects. At the beginning of this experiment, the only thing we had expected is just a low-noise amplification of spin waves, not at all the regeneration that is later observed.

\begin{figure}
\centering 
\includegraphics[width=0.98\columnwidth]{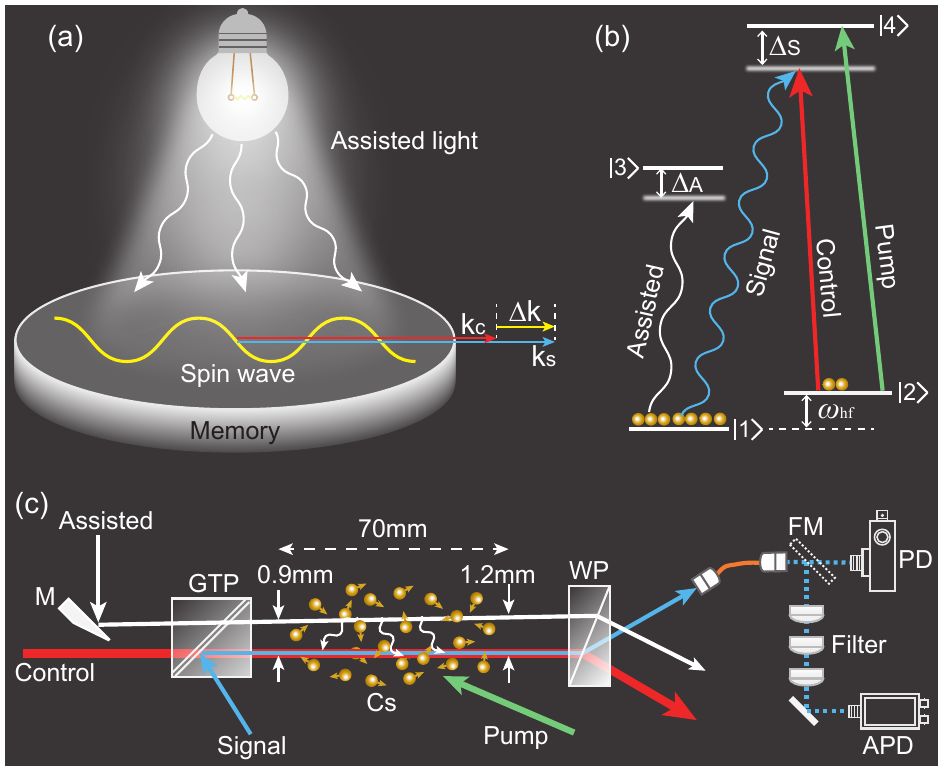}
\caption{{\textbf{Schematic diagram of the experiment.}} ({\bf a}) Conceptual graph of regenerating spin wave. We can view the spin wave as a living system with the ability to regenerate, and analogize the memory in which the spin wave is stored to a circular petri dish. A light source, such as a thermal light, is used to provide the energy for the regeneration of spin waves. ({\bf b}) Our protocol contains four main stages: 1st, initialize the atomic ensemble by a pump light, and almost all of the atoms populate on the the energy level $\left| 1\right\rangle$. 2nd, create an atomic spin wave (also known as atomic excitations) by signal pulse and control pulse, which is the map-in process of a Raman optical memory; 3rd, apply assisted photons to illuminates the spin wave, and provide energy for the regeneration of spin wave by Raman scattering. 4th, read out the spin wave as a signal pulse. The wave vector of the spin wave is $\Delta k$, and $|\Delta k|=\omega_{\rm hf}/c$. ({\bf c}) Experimental setup for demonstrating the regeneration of spin wave. The small white wavy lines in the caesium atomic gas are used to indicate assisted photons scattered from the assisted light. M: D-type Mirror; FM: flip mirror,  which allows alternative selection of detections via intensity or single photons; GTP: Glan-Taylor polarizer, for combining signal pulse and control pulse; WP: Wollaston prism; PD: photo diode; APD: avalanche photo-diode. There is a small angle around 4\,mRad between the assisted light and signal light, which is designed for reducing the noise leak from the assisted light.}
\label{fig1}
\end{figure}

Here, we demonstrate the regeneration of atomic spin wave in terms of its retrieval efficiency and lifetime, which depend on its spatial mode and coherence. We generate spin waves in warm alkali-metal atomic vapor by the write-in process of a Raman memory \cite{Reim2010, Guo2019}: use a strong control pulse (write pulse) to map a weak signal pulse onto atomic excitations, and the information about the intensity and phase of the signal pulse is coherently transferred into an atomic spin wave. The atomic spin wave is carried by a large number of atoms on the ground-state hyperfine energy levels $\left|1\right\rangle$ or $\left|2\right\rangle$. Then, an assisted light is incident into the atomic vapor, and its scattered photons named by assisted photons, coupling the transition $\left|1\right\rangle \to \left|3\right\rangle$ with a red detuning $\Delta_{\rm A}$, are utilized for amplifying the atomic spin wave, as is illustrated by the schematic diagram shown in Fig.\,\hyperref[fig1]{1}. There are three reasons for using such an assisted light. Firstly, the scattered assisted photons from the assisted light are not well orientated, and the direction and phase of the noisy spin wave generated by the assisted photons are also chaotic, which makes it difficult for the noisy spin wave to be read out efficiently by the control light (read pulses). In contrast, the interesting atomic spin wave that is coherently converted from the signal pulse has a preferred direction with a collective enhancement effect, and therefore can be read out efficiently \cite{Gorshkov2007}. Secondly, the wavelength of the assisted light (894.6 nm) differs from that of the signal light (852.3 nm) by several tens of nanometres. This make it easy to filter out the assisted light, while the assisted light is difficult to be coupled with the signal light and the control light, which suppresses the noise greatly. Thirdly, an atomic spin wave is able to regenerate by absorbing chaotic assisted photons, which can exactly verify that the atomic spin wave has self-organization characteristics similar to living organisms.

\begin{figure}[t!]
\centering
\includegraphics[width=1\columnwidth]{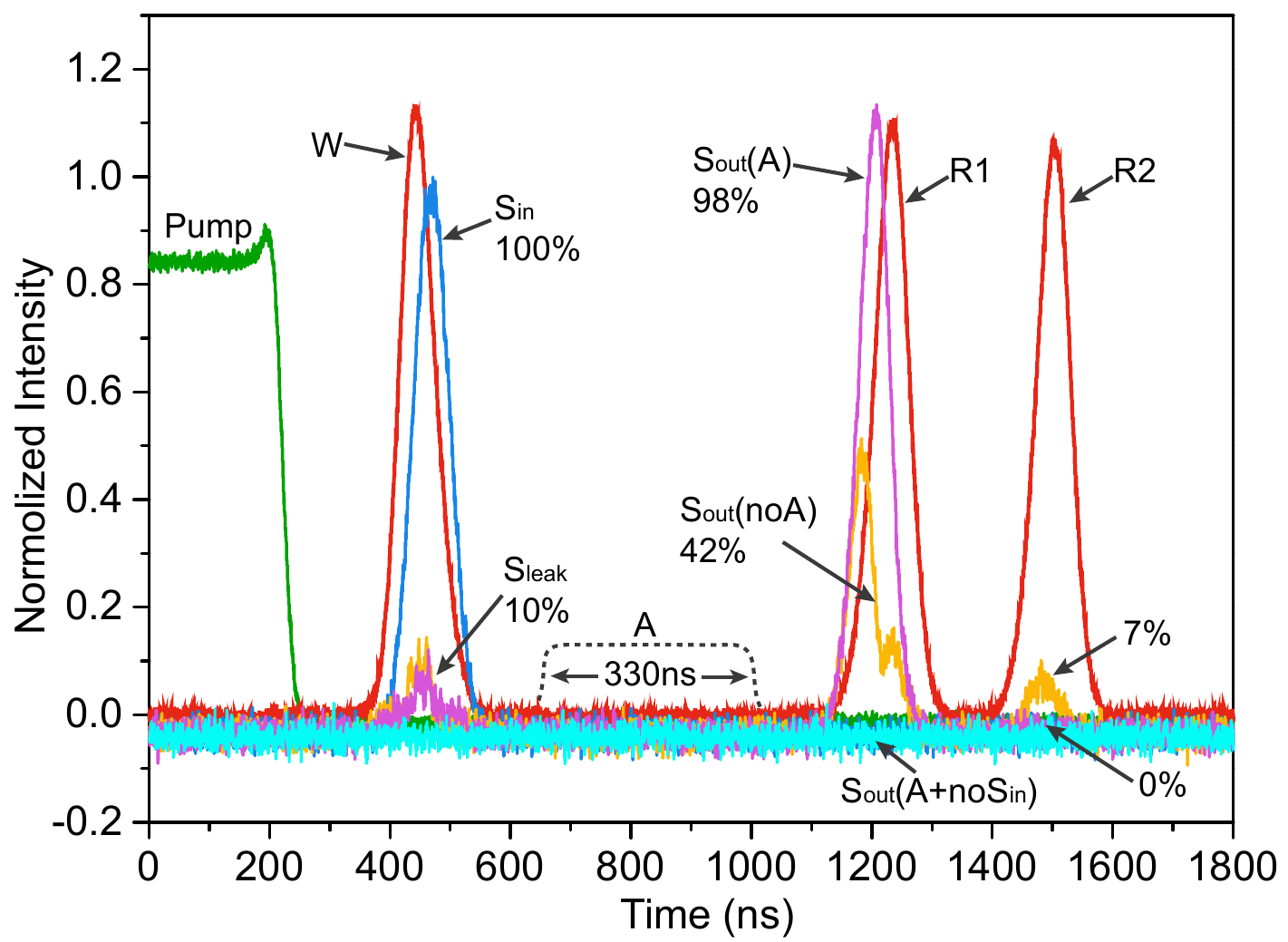}
\caption{\textbf{Retrieval efficiencies with or without assisted light.} The time delay between the write pulse and first read pulse is about 720\,ns. W: write pulse; R1: first read pulse;  R2: second read pulse; ${\rm S_{in}}$: input signal pulse with an energy of 13\,pJ, which is measured when the control pulses are turned off; A: assisted light; ${\rm S_{leak}}$: leaked signal; ${\rm S_{out}(noA)}$: retrieved signal pulse when no assisted light is applied; ${\rm S_{out}(A)}$: retrieved signal when a 330\,ns assisted light is turned on between W1 and R1; ${\rm S_{out}(A+noS_{in})}$: retrieved signal pulse when the assisted light is applied between W1 and R1, but no input signal pulse. Each efficiency value is calculated by dividing the detected pulse energy with that of ${\rm S_{in}}$. Consequently, the actual retrieval efficiency of a successfully stored atomic spin wave is about 0.1 times higher than what is shown in the plot. The control pulses (W, R1, and R2) and pump light have been normalized to align with the overall data plot, and their heights in the plot do not accurately reflect their energy ratios compared to the signal pulse.}
\label{fig2}
\end{figure}

The waveforms of the control pulse (i.e., write pulse and read pulse) and the signal pulse that we utilized are both approximated as Gaussian pulses with a pulse width about 70 ns, as shown in Fig.\,\hyperref[fig2]{2}. Coupled with the write pulse (W), the input signal pulse (${\rm S_{in}}$) is absorbed by the atoms and coherently converted into an atomic spin wave. The small peaks around 450\,ns show that about 90\% (10\%) of the signal pulse is absorbed (leaked). In order to verify whether the first read pulse (R1) can completely read out the stored spin wave, i.e., whether the retrieval efficiency is close to 100\%, we introduce another read pulse (R2) after R1. In the absence of the assisted light, the retrieval efficiency of R1 is limited to 42\%, and R2 continues to read out the signal pulse with a retrieval efficiency 7\%, depicted by the small yellow peak near 1500 ns, indicating that the R1 cannot completely read out the stored spin wave. On the contrary, when a 330\,ns assisted light is present between pulses W and R1, the observed retrieval efficiency is up to 98\% larger than the write-in efficiency 90\% due to the amplification. And R2 almost cannot read out any signal light (see the purple line around 1500/,ns), indicating that the spin wave has been completely read out by R1. These two cases are in sharp contrast in terms of retrieval efficiencies. In addition, we note that the pulse shape of ${\rm S_{out}(A)}$ is more like a Gaussian pulse with a regular shape than ${\rm S_{out}(noA)}$, see the purple line and yellow line around 1200\,ns.

Such results suggest that the atomic spin wave experiences not only a simple amplification, but also a life-like regeneration process, in which the spin-wave decay due to thermal motions of atoms is compensated, and the spatial mode of atomic spin wave is optimized, enabling a high retrieval efficiency. Note that the retrieval efficiency is independent of read pulse shape if the read pulse energy is high enough, but depends on the spatial mode of spin wave, as is suggested by Alexey V. Gorshkov {\it et al}. \cite{Gorshkov2007}. Yet, there is a question that needs futher efforts to explore: It is interesting to explain how a regeneration process used here enables the spin wave to evolve in a manner corresponding to a high retrieval efficiency. An intuitive interpretation is: When some atoms exit the interaction region or move a distance comparable to the wavelength of spin wave, the waves they carry undergo rapid decoherence, and therefore cannot be effectively read out. Conversely, those atoms entering the interaction region from outside will be incorporated in correct phase into the group of atoms carrying the right spin wave, providing fresh atoms on energy level $\left|1\right\rangle$ for regenerating the spin wave. It might be pointed out that, if the distributions of atoms and assisted photons are statistically isotropic, the central position of the atomic spin wave will not be biased in a particular spatial direction.

\begin{figure}[t]
\centering 
\includegraphics[width=0.96\columnwidth]{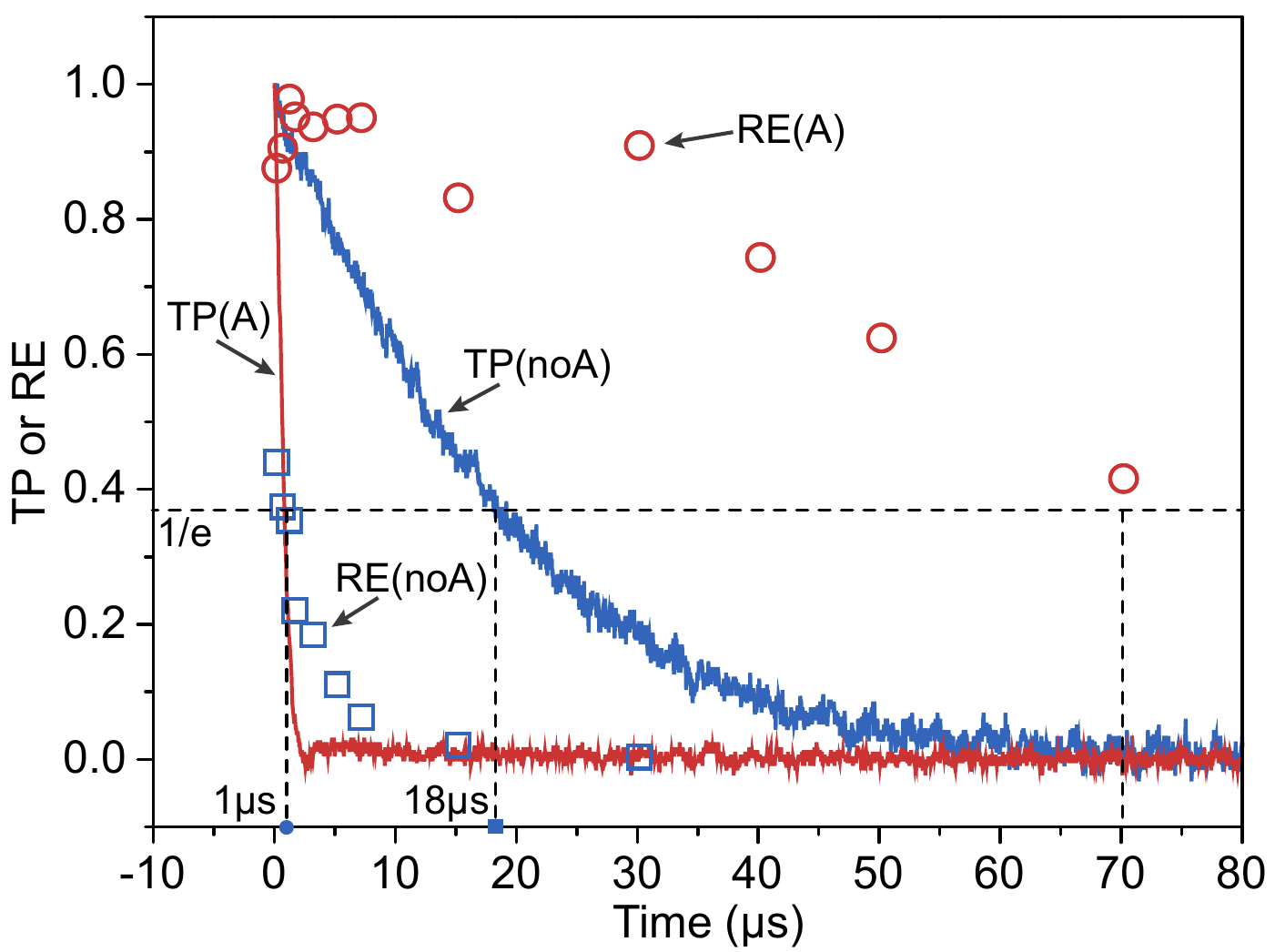}
\caption{\textbf{Transmission probability (TP) for characterizing the atomic-hyperfine-polarization lifetime, while retrieval efficiency (RE) for characterizing the spin-wave coherence lifetime.} A: an assisted light is applied to illuminate the stored spin wave during the interval between the write pulse and first read pulse; noA: there is no assisted light. For measuring the blue curve TP(noA), we utilize a vertically polarized probe light with a power of 5\,$\mu$W, and a horizontally polarized pump light with a power of 23\,mW. The probe light and pump light are resonant with the transition $\left | 2 \right \rangle \to  \left | 4 \right \rangle$, and are counter-propagating in the atomic ensemble. When the pump light is suddenly shut off at 0\,$\mu$s, the hyperfine relaxation which takes place during the interval of darkness causes the atomic ensemble to become more opaque to the probe light at a rate which is determined by the relaxation lifetime. Actually, this is Franzen’s classic technique of `relaxation in the dark' \cite{Franzen1959, Graf2005}. For measuring the red curve TP(A), we turn on the assisted light immediately after turning off the pump light, and observe an instantaneous decreasing of the transmission probability of the probe light. In the absence of assisted light, the $1/e$ lifetime of atomic hyperfine polarization is 18\,$\mu$s marked by a filled blue square, while the coherence lifetime is approximately 1\,$\mu$s marked by a filled blue circle. Error bars and theoretical curves are omitted as this study primarily aims to qualitatively demonstrate the lifetime enhancement through spin-wave regeneration, rather than provide a precise quantitative description. The fluctuations observed in the red circles (RE(A)) are attributed to experimental setup instabilities, such as the unlocked frequency and intensity of the assisted light during this experiment.
 }  
\label{fig3}
\end{figure}

Furthermore, we characterize the noise of the regeneration process. When there is no input signal pulse but the assisted light is present, neither R1 nor R2 can read out an observable signal, therefore the noise introduced by the control pulses and assisted light is negligible, see the cyan line ${\rm S_{out}(A+noS_{in})}$ around 1200\,ns and 1500\,ns in Fig.\,\hyperref[fig2]{2}. Note that, the cyan line is the intensity of retrieved signal pulse detected by a photo-diode detector (PD), whose sensitivity is not suitable for measuring a light pulse with an energy smaller than 0.5 pJ. To verify an accurate noise level, we flip the FM shown in Fig.\,\hyperref[fig1]{1(c)} and direct the retrieved photons to the frequency filter, and detect the ${\rm S_{out}(A+noS_{in})}$ by a single-photon detector (APD) \cite{Liu2022}. The detected average number (raw data) of noise photons introduced by a 330\,ns assisted light is 0.012, where the error bar is ignored due to a large total counts around 12000 by accumulating the data of $10^6$ trials. By dividing the value 0.012 with the total detection efficiency (7\%) of the signal photon, we obtain the intrinsic noise, 0.17 photons. Compared to the energy of several pJ added to the desired signal ${\rm S_{out}(A)}$ through spin-wave regeneration process, 0.17 photons are really negligible.

In Fig.\,\hyperref[fig3]{3}, the atomic hyperfine-polarization lifetime and the spin-wave coherence lifetime are characterized. For easy understanding, one can view the atomic hyperfine-polarization lifetime as the population lifetime of atoms on the energy level $\left|1\right\rangle$. By utilizing an assisted light between write pulse and read pulse, the coherence lifetime of spin wave is up to 70\,$\mu$s, and is longer than the atomic hyperfine-polarization lifetime 18\,$\mu$s. It is worth noting that the red curve TP(A) is much steeper than TP(noA), which we attribute to the assisted light that pump the atoms from energy level $\left|1\right\rangle$ onto $\left|2\right\rangle$. This pump results in fewer atoms populating on the level $\left|1\right\rangle$, and leads to insufficient channels $\left|1\right\rangle \to \left|3\right\rangle \to \left|2\right\rangle$ for assisted light to continuously energize the spin wave. This is the reason why our demonstration of spin-wave regeneration is limited to a lifetime smaller than 100\,$\mu$s at the present.

There are two possible approaches to resolve or mitigate this depolarization problem due to assisted light: Firstly, optimize the generation pattern of assisted photon, for example, illuminate the spin wave directly by a weak thermal light with chaotic directions, rather than the one scattered from an intense assisted coherent light utilized in this article. Secondly, the atoms transferred into the state $\left|2\right\rangle$ can be pumped back into the initial state $\left|1\right\rangle$ promptly by introducing an additional pump light that does not spatially overlap with the interesting spin wave. Due to thermal motion, the atoms return to the state $\left|1\right\rangle$ will rapidly spread throughout the whole atomic cell, including the region where the spin wave locates. This ensures that the spin wave has a capability of sustainable regeneration. 

In summary, we have demonstrated that the lifetime and retrieval efficiency of atomic spin wave can be significantly enhanced by a life-like regeneration process, while the noise is effectively suppressed. Particularly, the observed intrinsic noise, in the order of 0.1, suggests that such a regeneration scheme may be harnessed in the realms of quantum information and precision measurements, where long lifetime, high efficiency and low noise are key features. Our results are going to trigger further works, including investigations into whether the regeneration mechanism used here can position the storage media with severe spin-wave decoherence, such as room-temperature atomic vapor and colour centers, as promising candidates for long-life and high-efficiency quantum memory.

Among challenges faced in room-temperature systems, we emphasize that noise reduction is paramount important: efforts are needed to further suppress intrinsic noise introduced by the assisted light. Additionally, another notable source of noise is the well-known four-wave mixing (FWM) noise due to control pulses in atomic ensemble-based $\Lambda$-type quantum memories. In our experiment, the intrinsic FWM noise is around 0.8 photons, which hinders the demonstration of a non-classical single-photon memory potentially enhanced by our regeneration scheme. It is worth mentioning that some solutions have been proposed, such as magic detuning \cite{Dideriksen2021}, cavity enhancement \cite{Nunn2017} and destructive interference \cite{Thomas2019}, to mitigate the challenge of FWM noise. These are of great significance in advancing low-noise room-temperature quantum memory, showcasing innovative and forward-looking results.

\begin{acknowledgements}
The authors thank Jian-Wei Pan for helpful discussions. This research is supported by the National Key R\&D Program of China (Grants No. 2019YFA0308703, No. 2019YFA0706302, and No. 2017YFA0303700); National Natural Science Foundation of China (NSFC) (Grants No. 62235012, No. 11904299, No. 61734005, No. 11761141014, and No. 11690033, No. 12104299, and No. 12304342);  Innovation Program for Quantum Science and Technology (Grants No. 2021ZD0301500, and No. 2021ZD0300700); Science and Technology Commission of Shanghai Municipality (STCSM) (Grants No. 20JC1416300, No. 2019SHZDZX01, No.21ZR1432800, and No. 22QA1404600); Shanghai Municipal Education Commission (SMEC) (Grants No. 2017-01-07-00-02-E00049); China Postdoctoral Science Foundation (Grants No. 2020M671091, No. 2021M692094, No. 2022T150415). X.-M.J. acknowledges additional support from a Shanghai talent program and support from Zhiyuan Innovative Research Center of Shanghai Jiao Tong University.\\
\end{acknowledgements}

\noindent{\Large\bf Methods}  

\noindent \textbf{Details of the experimental setup.} The $^{133}$Cs atoms are sealed in a glass cell with a length of 75\,mm. The glass cell is packed in a three-layer magnetic shielding and is heated up to 72\,$^{\circ}$C. The lower two energy states are $\left | 1 \right \rangle \left( 6S_{1/2}, F=3\right)$ and $\left | 2 \right \rangle\left( 6S_{1/2}, F=4\right)$, {\it i.e.} the hyperfine ground states of $^{133}$Cs, the excited state $\left | 3 \right \rangle$ is the $ 6P_{1/2}$ manifold. and the excited state $\left | 4 \right \rangle$ is the $ 6P_{3/2}$ manifold. The initial state refers to the state that almost all atoms populate on the energy level $\left | 1 \right \rangle$. The pump light for preparing atomic initial state comes from an external cavity diode laser, and is resonant with the transition $\left( 6S_{1/2}, F=4 \to  6P_{3/2} \right)$. The pump light is turned on at least 1\,$\mu{\rm s}$ and is turned off about 100\,ns earlier than the arrival of the first control pulse. 

The control pulses come from a distributed Bragg reflector (DBR) laser. The signal pulses come from a distributed Feedback (DFB) laser. For each pulse, a fast electro-optic modulator (EOM) triggered by electronic pulses from an arbitrary waveform generator is used to chop the laser to short pulses with tunable amplitudes. The generated control pulses are fed into a tapered amplifier (TA) to boost their power. In order to eliminate the spontaneous emission from the TA, we employ a ruled diffraction grating to spread beam out and spatially pick the stimulated radiation with irises. Before entering the cells, the control pulses are horizontally polarized by a Glan-Taylor polarizer, and the signal pulses are vertically polarized. In this letter, the control pulse width is 70\,ns, and the beam waist in the glass cell is about 300\,$\mu$m. The control pulse energy is set to around 2\,nJ. The frequency of control pulses are locked to that of the pump light, and are red detuned ${2.9\,{\rm GHz}}$ from the transition $\left( 6S_{1/2}, F=4 \to 6P_{3/2}, F=4 \right) $. The signal pulse width is around 70\,ns, and the beam waist in the glass cell is about 240\,$\mu$m. The signal pulse energy is around 13\,pJ for the intensity detection. The frequency of signal pulses are locked to that of the pump light, and are red detuned ${2.9\,{\rm GHz}}$ from the transition $\left( 6S_{1/2}, F=3 \to 6P_{3/2}, F=4 \right) $. The assisted light (894.6\,nm) is red detuned around ${0.8\,{\rm GHz}}$ from the transition $\left( 6S_{1/2}, F=3 \to 6P_{1/2}, F=3 \right) $, and its power is set to around 10\,mW. The beam waist of assisted light is 190\,um.

Here, we use a Wollaston prism (shown in Fig.\,\hyperref[fig1]{1}) to basically separate the retrieved signal pulse from the control pulses and assisted light. For single-photon experiment, there are still too much noise just after the Wollaston prism due to the leakage or scattering of control pulses. In our experiment, we use a home-made frequency filter (cascaded cavities) to filter out noise photons. The peak transmission of each cavity is higher than 90\%. The peak transmission of the whole filter is about 70\%, and the extinction ratio is about $10^{7}:1$.  The full width at half maximum of the total transmission window is 380\,MHz. After passing through the cascaded cavities, the signal photons are detected by an APD with a quantum efficiency around 50\%. In addition, there some fiber collimators for collecting or emitting the signal photons, which result in a low total detection efficiency about 7\%.

\hspace*{\fill}\\
\noindent \textbf{Noise suppression through atomic collective effects and momentum conservation.} Assume that there are $n_{{\rm{E}}}(0)$ atomic excitations at initial time $t=0$. When an assisted light is incident onto the atomic ensemble, an amplification process for atomic excitations happens, which is induced by the photons scattered from the assisted light. After some time $t$, the average number of collective atomic excitation is
\begin{equation}\label{eqZaisheng}
 \bar n_{\rm{E}} (t) =  n_{{\rm{E}}}(0) \cosh ^2 (\kappa t) + \sinh ^2 (\kappa t),
\end{equation}
where $\kappa$ is a coupling coefficient \cite{Louisell1961, Chen2009}. The first term is the amplified atomic excitations. It is the amplification or gain process that regenerates the decayed atomic excitations. The last term corresponds to noise due to spontaneous Raman scattering of assisted photons, and should be suppressed or filtered out. The variance of $n_{\rm{E}} (t)$ can be written as 
\begin{equation}\label{eqFangcha}
 \left(\Delta n_{\rm{E}}  \right)^2 =\sinh^2(2\kappa t)\left[ 1+ n_{{\rm{E}}}(0) \right]/4 , \\ 
\end{equation}

From the above two equations, we can see that it is $\rm sinh(\cdots)$ that introduces noise and fluctuations to the interesting spin wave. A possible way to get rid of these noise and fluctuations is the destructive interference due to a large number of atoms interacting with the assisted photon \cite{Scully2006}:
\begin{equation}\label{destructive}
 \sum_{j}{e^{i(\vec{k}_{\rm A}-\vec{k}_{\rm RA})\cdot \vec{r}_j}} \propto \delta^3(\vec{k}_{\rm A}-\vec{k}_{\rm RA})=0,
\end{equation}
where $\vec{k}_{\rm A}$ is the wave vector of assisted photon, and $\vec{k}_{\rm RA}$ is the wave vector of the spontaneous Raman-scattered photon excited by assisted photon. $ \vec{r}_j$ is the position of the $j$th atom. Note that $|\vec{k}_{\rm A}|-|\vec{k}_{\rm RA}|= \omega_{\rm hf}/c \neq 0$. According to (\ref{destructive}), there is a destructive interference between different atoms, and thus the spontaneous Raman scattering distributes over all possible atomic modes, and has no preferred direction. In a word, the noisy spin wave due to assisted photons has no definite direction.

On the contrary, the interesting spin wave generated by signal light and control light has a definite direction along the one of $(\vec{k}_{\rm S}-\vec{k}_{\rm C})$, where $\vec{k}_{\rm S}$ and $\vec{k}_{\rm C}$ are the wave vectors of signal light and control light, respectively. During the read process, the noisy spin wave from spontaneous Raman scattering can not be efficiently retrieved into the desired spatial mode due to the momentum mismatch:
\begin{equation}
\vec{k}_{\rm A}-\vec{k}_{\rm RA}\neq\vec{k}_{\rm S}-\vec{k}_{\rm C}.
\end{equation}
This is the reason why the noise due to spontaneous Raman scattering can be removed or filtered when a single mode fiber is used to collect the retrieved signal light.

It is worth mentioning that, there is a distance about 1\,mm between the assisted laser and signal light. By scattering the assisted laser, we obtain the assisted photons that illuminate the interesting spin wave. The assisted photon arriving at the interesting spin wave has a direction much different from the one of signal light, since the assisted photons with a direction parallel to signal light can not arrive at the interesting spin wave. Therefore, even if the spatial mode of assisted photons is a Gaussian beam, and the noise spin wave is directed along the Gaussian beam due to collective enhancement \cite{Duan2001, Duan2002}, its direction is much different from the one of the interesting spin wave, and thus can not be effectively read out and collected.





\end{document}